\documentclass[conference]{IEEEtran}
\IEEEoverridecommandlockouts
\usepackage[T1]{fontenc}
\usepackage{graphicx}
\usepackage{url}
\usepackage{multicol}
\usepackage{algorithmic}
\usepackage[perpage, symbol*]{footmisc}
 \usepackage{amsmath}
 \usepackage{amsfonts}
 \usepackage{setspace}
 \usepackage{subfigure}
 \usepackage{algorithm}  
  \usepackage{multirow}
  \usepackage{threeparttable}
  \usepackage{xcolor}
\newcommand{\reffig}[1]{Fig.\ref{#1}}

\begin{document}
\title{Transfer Learning in Financial Time Series with Gramian Angular Field}

\author{
    \IEEEauthorblockN{1\textsuperscript{st} Hou-Wan Long}
    \IEEEauthorblockA{
        \textit{The Chinese University of Hong Kong, Hong Kong}\\
        houwanlong@link.cuhk.edu.hk}
    \and
    \hspace{-1em} 
    \IEEEauthorblockN{2\textsuperscript{nd} On-In Ho}
    \IEEEauthorblockA{
        \textit{University of Macau, Macao}\\
        MC15410@um.edu.mo}
    \and
    \hspace{-4em} 
    \IEEEauthorblockN{3\textsuperscript{rd} Qi-Qiao He}
    \IEEEauthorblockA{
        \textit{Foshan University, China}\\
        hesammul@gmail.com}
    \and
    \hspace{-1em} 
    \IEEEauthorblockN{4\textsuperscript{th} Yain-Whar Si}
    \IEEEauthorblockA{
        \textit{University of Macau, Macao}\\
        fstasp@um.edu.mo}
}

\maketitle              
\begin{abstract}
In financial analysis, time series modeling is often hampered by data scarcity, limiting neural network models' ability to generalize. Transfer learning mitigates this by leveraging data from similar domains, but selecting appropriate source domains is crucial to avoid negative transfer. This study enhances source domain selection in transfer learning by introducing Gramian Angular Field (GAF) transformations to improve time series similarity functions. We evaluate a comprehensive range of baseline similarity functions, including both basic and state-of-the-art (SOTA) functions, and perform extensive experiments with Deep Neural Networks (DNN) and Long Short-Term Memory (LSTM) networks. The results demonstrate that GAF-based similarity functions significantly reduce prediction errors. Notably, Coral (GAF) for DNN and CMD (GAF) for LSTM consistently deliver superior performance, highlighting their effectiveness in complex financial environments.
\end{abstract}

\begin{IEEEkeywords}
    Time Series, Transfer Learning, Similarity Functions, Gramian Angular Field, Deep Learning
\end{IEEEkeywords}

\section{Introduction}
Transfer learning is a powerful and increasingly popular machine learning technique designed to enhance the performance of models in a target domain by leveraging knowledge acquired from a related source domain. This technique is particularly valuable in scenarios where the target domain suffers from data scarcity, which often hinders the development of robust models and increases the risk of overfitting. By effectively transferring information from a source domain that has a more abundant and rich dataset, transfer learning enhances model accuracy and generalization capabilities. This makes it an essential tool across various fields, especially in situations where labeled data is limited or expensive to obtain. The ability of transfer learning is evident in its widespread application across diverse areas, particularly in time series analysis. It has proven to be effective in domains such as fault diagnosis, human activity recognition (HAR), and brain-computer interfaces (BCI), where time-dependent patterns are crucial for accurate predictions. A systematic study by Weber et al. \cite{weber2021transfer} reviewed 223 publications and highlighted the extensive application of transfer learning across more than 40 different time series domains, underscoring its broad utility and impact.

However, the process of transferring knowledge is not without its challenges. One of the most significant risks associated with transfer learning is the phenomenon known as negative transfer. Negative transfer occurs when the inclusion of data from a source domain, instead of enhancing, actually deteriorates the model’s performance in the target domain. This issue, first identified by Rosenstein et al. \cite{rosenstein2005transfer}, has been a focal point of transfer learning research, prompting the development of various strategies to mitigate its effects. For instance, in time series classification, Fawaz et al. \cite{fawaz2018transfer} observed instances of negative transfer when using Convolutional Neural Networks (CNNs). To address this, researchers have proposed various sophisticated domain selection techniques to better align the source and target domains and reduce the likelihood of negative transfer \cite{yao2010boosting}. These techniques aim to ensure that the source domain's data is sufficiently relevant and beneficial to the target domain, thereby maximizing the effectiveness of the transfer learning process.

A crucial aspect of mitigating negative transfer and ensuring effective knowledge transfer is the selection of an appropriate source domain, which relies heavily on similarity functions. Similarity functions play a key role in measuring how closely related the source and target domains are. They do so by quantifying the distance between distributions or feature representations of the two domains. This measurement is often referred to as a distance measure. Distance measures, such as Euclidean distance, Pearson correlation, Maximum Mean Discrepancy (MMD) \cite{chen2018cross-position}, Central Moment Discrepancy (CMD) \cite{zellinger2017central}, Correlation Alignment (CORAL) \cite{sun2016correlation}, Dynamic Time Warping (DTW), Time Warp Edit Distance (TWED) \cite{marteau2009twed}, and Wasserstein distance \cite{courty2017optimal}, serve as the mathematical foundation for similarity functions. These measures calculate the degree of difference or similarity between the domains, guiding the transfer learning process by indicating the most relevant and beneficial source domain for the target task. For instance, Euclidean distance calculates the straight-line distance between two points in a feature space, while DTW aligns sequences that may vary in time or speed, making it particularly useful for time series data. Similarly, MMD and CMD measure discrepancies in the statistical distributions of the source and target domains.

Despite the wide variety of similarity functions and distance measures currently available, their effectiveness in the context of transfer learning—especially for time series data—remains underexplored. The selection of an appropriate similarity function is crucial for minimizing negative transfer and enhancing model performance, yet there has been insufficient research into how these functions perform across different transfer learning scenarios. Specifically, in time series analysis, where data is inherently complex and temporally dependent, there is a notable gap in comprehensive evaluations of these methods. Given that time series data often contain rich temporal and angular relationships, we are motivated to explore whether enhancing these similarity functions with Gramian Angular Field (GAF) \cite{wang2015imaging} transformations could significantly improve their accuracy and robustness. To address these gaps, we not only evaluate the existing baseline similarity functions, including both basic and SOTA functions, but also perform extensive experiments to assess how these functions perform when enhanced with GAF transformations.

Gramian Angular Field (GAF) transformation is a technique that transforms time series data into two-dimensional images (see \reffig{transform}), capturing both the temporal structure and angular relationships within the data. GAF has shown potential in preserving the intrinsic patterns of time series data, making it a valuable tool for enhancing similarity functions. By converting time series data into a form that can be processed similarly to images, GAF allows for the extraction of richer, more informative features that can improve the accuracy of domain distance measurements.

Our contributions are threefold:
\begin{itemize}
\item[(i)] We evaluate baseline similarity functions for source dataset selection in both single-source and multi-source transfer learning scenarios.
\item[(ii)] We introduce a GAF-based method for measuring domain distance and demonstrate that GAF scaling significantly improves performance compared to unscaled similarity functions.
\item[(iii)] We provide a detailed analysis of the reasons behind the superior performance of GAF-based methods, offering deeper insights into their comparative advantages over traditional similarity functions.
\end{itemize}

The paper is organized as follows: Section 2 reviews related literature, Section 3 introduces the GAF-based similarity functions, Section 4 evaluates the baseline similarity functions for transfer learning, Section 5 is the experiment section, and Section 6 concludes the paper.

\section{Related Work}
This section reviews the use of deep learning models and advanced similarity functions to enhance transfer learning in time series.

\subsection{Time Series Problems and Transfer Learning Approaches}
Time series problems supported by transfer learning include classification, regression, and clustering, as noted by Weber et al. \cite{weber2021transfer}. Anomaly detection and time forecasting fall under classification and regression. In their study, 223 publications were included, with 164 focused on time series classification. Transfer learning in time series primarily uses deep learning models, with four main approaches: model-based, feature-based, instance-based, and hybrid. Model-based transfer learning is the most common, utilizing CNN, RNN, and LSTM models, with combinations like CNN-RNN and CNN-LSTM also prevalent. Model-based techniques include pre-training and fine-tuning, partial freezing, and architecture modification. Domain-adversarial learning, dedicated model objectives, and ensemble-based transfer are also used. Transforming time series into GAF, proposed by Oates and Wang \cite{wang2015imaging}, is notable in transfer learning.


\begin{figure}[htb]
    \centering
    \includegraphics[width=8.5cm]{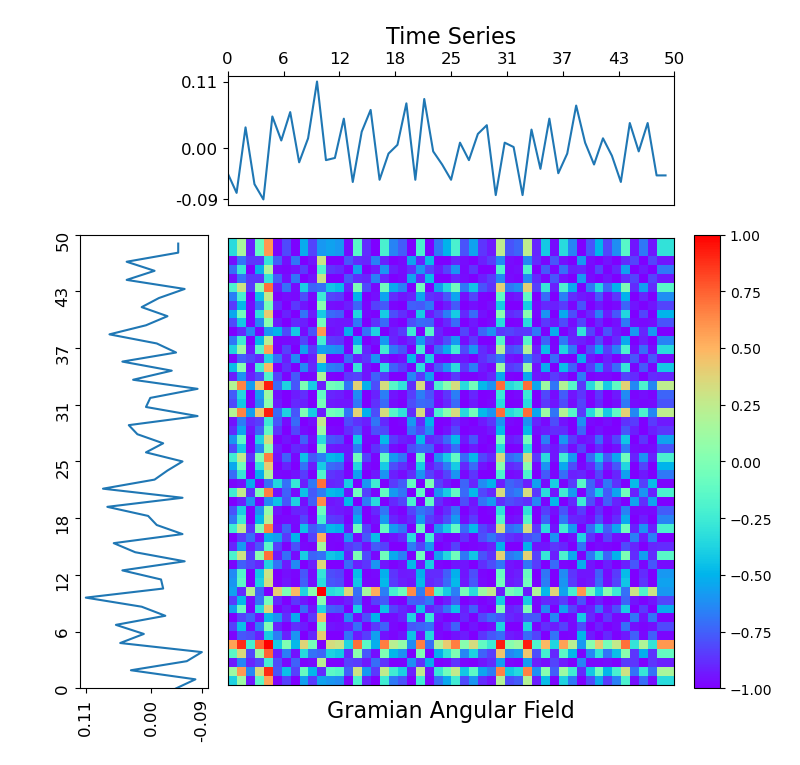}
    \caption{GAF Transformation on Time Series}
    \label{transform}
\end{figure}

\begin{figure*}[htb]
    \centering
    \includegraphics[width=\textwidth]{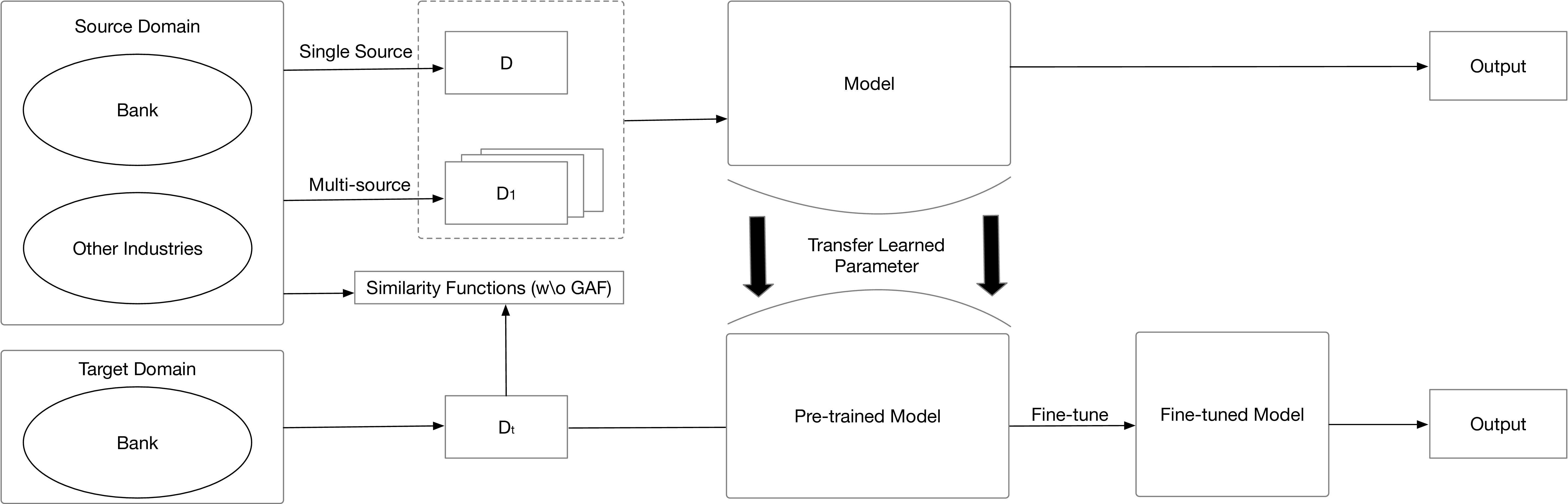}
    \caption{The Procedure of GAF-based Transfer Learning}
    \label{flow2}
\end{figure*}

\subsection{Distance Measures and Transfer Learning Techniques}
Ben-David et al. \cite{Ben-David2007} utilized A-distance, a type of distance measurement, to quantify the difference between domains in transfer learning. This approach laid the groundwork for the development of more advanced similarity functions. Chen et al. \cite{chen2018cross-position} built upon this by introducing a transfer learning method that uses fine-grained Maximum Mean Discrepancy (MMD) as a distance measure to assess domain discrepancies, serving as a similarity function that guides the transfer process. Wang et al. \cite{Wang2018} and Wang et al. \cite{Wang2020} further refined this approach by estimating a balance factor $\mu$, which adjusts the balance between marginal and conditional distribution adaptation, with A-distance being a key component in defining the similarity function.

In geometrical feature transformation methods, Sun et al. \cite{sun2016correlation} proposed CORrelation Alignment (CORAL), which employs second-order statistics as a similarity function to align the source and target domains. The core of CORAL is learning a transformation matrix that minimizes the distance between the second-order statistics of the source and target domains, thereby reducing distribution divergence. This method was extended to Deep CORAL \cite{Sun2016deep}, where CORAL loss replaced MMD loss, effectively acting as a new similarity function for aligning feature distributions. Wasserstein distance, another critical distance measure rooted in optimal transport theory, was applied in transfer learning by Courty et al. \cite{courty2017optimal}. They used it as a similarity function to learn feature transformations that reduce the marginal distribution distance, further extending this approach with Joint Distribution Optimal Transport (JDOT) to account for both marginal and conditional distributions \cite{Courty2017bb}.

In deep transfer learning, Tzeng et al. \cite{tzeng2014deep} introduced Deep Domain Confusion (DDC), where MMD was used as a distance measure to define a similarity function for aligning feature distributions in deep networks. Long et al. \cite{Long2015} advanced this with Deep Adaptation Networks (DAN), which replaced single-kernel MMD with multiple-kernel MMD, offering a more flexible similarity function for distribution adaptation. Zhu et al. \cite{Zhu2021} proposed Deep Subdomain Adaptation Networks (DSAN), introducing a local MMD distance that computes weighted MMD distances as a refined similarity function. Zellinger et al. \cite{zellinger2017central} proposed a new regularization method that directly minimizes discrepancies between domain-specific latent feature representations in the hidden activation space, introducing Central Moment Discrepancy (CMD) as a novel distance-based similarity function. This approach was tested on benchmark datasets for object recognition and sentiment analysis.

In the context of time series classification, Serrà and Arcos \cite{serra2014empirical} evaluated various distance measures used as similarity functions, including Euclidean distance, Fourier Coefficients (FC), auto-regression model-based (AR) measures, DTW, Edit Distance on Real Sequences (EDR), Time Warp Edit Distance (TWED), and Minimum Jump Costs dissimilarity (MJC). Their study concluded that TWED, as a similarity function, consistently outperformed others. No significant differences were found among DTW, EDR, and MJC, while Euclidean distance, another basic similarity function, performed worse than these methods. FC and AR measures, also acting as similarity functions, were ranked below Euclidean distance in performance.

\section{Transfer Learning Based on Gramian Angular Field (GAF)}

There are different variations of transfer learning, including single source transfer learning and multi-source transfer learning.

In the case of single source transfer learning, the procedure starts with the training of a model on a single source dataset, denoted as $D$. During this stage, the model learns the patterns, features, and knowledge inherent to this specific domain. Once the model has been trained, the learned parameters are transferred to the next stage, where the model undergoes fine-tuning using data from the target domain $D_t$, which, in this scenario, belongs to the bank sector. The fine-tuning process is essential as it allows the model to adjust and adapt the pre-learned knowledge to better fit the unique characteristics of the target domain's data. After fine-tuning, the model is tested on the test set of $D_t$ to evaluate its performance and to ensure that the model generalizes well to the new domain, ultimately producing the final output.

To ensure the selection of the most relevant source domain for transfer learning, a set of similarity functions is applied. The selection algorithm iterates over each similarity function $F$ in the set of similarity functions. For each function, the algorithm calculates the similarity between each source dataset $D_s$ in the set of possible sources $S$ and the target dataset $D_t$. If the GAF transformation is being used, both $D_s$ and $D_t$ are transformed before calculating the similarity. These calculated similarity values are stored in a list. If the similarity function $F$ is Pearson correlation, the algorithm selects the source dataset with the maximum similarity value. For other similarity functions, the source dataset with the minimum similarity value is selected. The chosen source dataset $D$ is then used for the initial training phase of the transfer learning model. The pseudo-code for this process is shown in Algorithm \ref{single-source-select}.
\begin{algorithm}[htp!]
\scriptsize
\begin{algorithmic}
\FOR{F in similarityfunctions}
    \STATE values = [ ]
    \FOR{$D_s$ in $S$}
        \IF{use\_GAF\_transform}
            \STATE $D_s \gets GAF(D_s)$
            \STATE $D_t \gets GAF(D_t)$
        \ENDIF
        \STATE value = $F(D_s, D_t)$
        \STATE values.append(value)
    \ENDFOR
    \IF{F == Pearson}
        \STATE $i = \text{argmax(values)}$
    \ELSE
        \STATE $i = \text{argmin(values)}$
    \ENDIF
    \STATE $D = S[i]$
\ENDFOR
\end{algorithmic}
\caption{Selecting Source Domains for Single-source Transfer Learning}
\label{single-source-select}
\end{algorithm}

For multi-source transfer learning, the process involves training the model on multiple datasets, denoted as 
$D_1$, $D_2$, and potentially others, which may come from different source domains like bank and other industries. This approach is designed to integrate diverse knowledge from various domains, with the aim of enhancing the model's robustness and generalization capabilities. The model, having been trained on these multiple sources, is then fine-tuned on the target domain $D_t$, similar to the single-source approach. The fine-tuning process refines the model's performance specifically for the target data, leveraging the combined knowledge from the multiple sources. Subsequently, the fine-tuned model is tested on the test set of $D_t$, and the final output is generated, ensuring that the model performs well on the new domain.

The selection of source domains in the multi-source scenario follows a more complex algorithm. Initially, the similarity function is applied to identify the first most relevant source dataset, denoted as $D_1$, from the set of possible sources $S$. If the GAF transformation is being used, both $D_s$ and $D_t$ are transformed before calculating the similarity. After selecting $D_1$, it is removed from $S$, and the similarity function is then reapplied to the remaining datasets in the modified set $S'$. This process identifies the second most relevant source dataset, denoted as $D_2$. The algorithm ensures that the selected datasets $D_1$ and $D_2$ are the most appropriate for training the model in a multi-source transfer learning setting. The pseudo-code for this multi-source selection process is shown in Algorithm \ref{multi-source-select}.
\begin{algorithm}[htp]
\scriptsize
\begin{algorithmic}
\FOR{F in similarityfunctions}
    \STATE values = [ ]
    \FOR{$D_s$ in $S$}
        \IF{use\_GAF\_transform}
            \STATE $D_s \gets GAF(D_s)$
            \STATE $D_t \gets GAF(D_t)$
        \ENDIF
        \STATE value = $F(D_s, D_t)$
        \STATE values.append(value)
    \ENDFOR
    \IF{F == Pearson}
        \STATE $i = \text{argmax(values)}$
    \ELSE
        \STATE $i = \text{argmin(values)}$
    \ENDIF
    \STATE $D_1 = S[i]$
    \STATE $S' = S.remove(D_1)$
    \STATE values = [ ]
    \FOR{$D_s$ in $S'$}
        \IF{use\_GAF\_transform}
            \STATE $D_s \gets GAF(D_s)$
            \STATE $D_t \gets GAF(D_t)$
        \ENDIF
        \STATE value = $F(D_s, D_t)$
        \STATE values.append(value)
    \ENDFOR
    \IF{F == Pearson}
        \STATE $i = \text{argmax(values)}$
    \ELSE
        \STATE $i = \text{argmin(values)}$
    \ENDIF
    \STATE $D_2 = S'[i]$
\ENDFOR
\end{algorithmic}
\caption{Selecting Source Domains for Multi-source Transfer Learning}
\label{multi-source-select}
\end{algorithm}

A critical component in our proposed transfer learning process is the use of GAF-based similarity functions to measure the distance between the source domains and the target domain. This approach addresses a key limitation of the baseline similarity functions discussed in Section \ref{exist}, which operate on 1-dimensional time series data and may not fully capture the complexity of the data. By transforming the data into a more representative form, these similarity functions can more effectively compare the source domain datasets with the target domain dataset $D_t$. The primary purpose of these functions is to aid in selecting the most appropriate source domains, thereby optimizing the transfer learning process and minimizing the risk of negative transfer. By accurately measuring the similarity between domains, these functions ensure that the knowledge being transferred is both beneficial and relevant to the target task.

\subsection{Turning 1-d Time Series to 2-d Gramian Angular Field}
There are three steps in transforming time series to GAF. The first step is to scale the time series to [-1,1] using:
\begin{equation}
\tilde{x}_i = \frac{(x_i - \max(X)) + (x_i - \min(X))}{\max(X) - \min(X)}
\end{equation}
$x_i$ represents an individual data point in the time series. $X$ denotes the entire time series dataset. The equation scales each data point to a range of [-1,1] by adjusting it relative to the maximum and minimum values in the entire dataset $X$.

Next, the scaled time series is transformed to polar coordinates:
\begin{equation}
\left\{
\begin{array}{ll}
\phi = \arccos(\tilde{x}_i), & -1 \leq \tilde{x}_i \leq 1, \tilde{x}_i \in \tilde{X} \\
r = \frac{t_i}{N}, & t_i \in \mathbb{N}
\end{array}
\right.
\end{equation}
where $\phi$ is the angle and $r$ is the radius.

Finally, the polar coordinates are converted to Gramian Angular Summation Field (GASF) image as shown in \reffig{transform}:
\begin{equation}
GASF = \left[ \cos(\phi_i + \phi_j) \right]
\end{equation}
where GASF represents the pixel value at the $(i, j)$ position in the GASF image. 

According to Oates and Wang \cite{wang2020transfer}, GAFs preserve temporal dependencies. As we move from the top-left to the bottom of GAFs, time increases, capturing temporal correlations effectively. GASF is a specific type of GAF that focuses on the summation of angular values. In this study, we generally refer to these transformations as GAF to encompass both GASF and other potential variations.

\subsection{Similarity Functions Applied on GAF}
Pearson, DTW, and TWED are designed for 1-dimensional series and are excluded for being applied on GAF. MMD is also excluded due to high memory consumption. Thus, the similarity functions applied on GAF include (i) Euclidean distance, (ii) Coral, (iii) CMD, and (iv) Wasserstein distance.

Additionally, three image similarity functions are applied on GAF. Before applying these, GAFs are scaled from [-1,1] to [0,255]:
\begin{equation}
GAF_{\text{scaled}} = GAF \times 127.5 + 127.5
\end{equation}

The image similarity functions include Adapted Rand error (ARE), Peak Signal Noise Ratio (PSNR), and Structural Similarity (SSIM). Skimage metrics module \cite{wang2004image} is used for these calculations.

Adapted Rand error is calculated as:
\begin{equation}
ARE=1 - \frac{\sum_{ij} p_{ij}^2}{\alpha \sum_k s_k^2 + (1 - \alpha) \sum_k t_k^2}
\end{equation}
where \( p_{ij} \) is the probability that the label of a pixel in the test image matches the true image, \( t_k \) and \( s_k \) are the probabilities of pixel labels in the true and test images, respectively, and \( \alpha \) is set to 0.5.

Structural similarity (SSIM) \cite{wang2004image} is defined as:
\begin{equation}
SSIM(x, y) = \frac{(2\mu_x\mu_y + c_1)(2\sigma_{xy} + c_2)}{(\mu_x^2 + \mu_y^2 + c_1)(\sigma_x^2 + \sigma_y^2 + c_2)}
\end{equation}
where \( c_1 = (k_1L)^2 \) and \( c_2 = (k_2L)^2 \). \( L \) is the dynamic range of pixel values, set to 255, and \( k_1 \) and \( k_2 \) are small constants. $x$ and $y$ represent the two images (or signals) being compared for structural similarity. $\mu_x$ represents the mean of the variable $x$, while $\mu_y$ represents the mean of the variable $y$. Similarly, $\sigma_x^2$ denotes the variance of $x$, $\sigma_y^2$ denotes the variance of $y$, and $\sigma_{xy}$ represents the covariance between $x$ and $y$.

Peak Signal Noise Ratio (PSNR) measures the ratio between the maximum possible power of a signal and the power of noise affecting image quality:
\begin{equation}
PSNR = 10 \cdot \log_{10} \left( \frac{MAX_I^2}{MSE} \right)
\end{equation}
where \( MSE \) is the mean squared error between two images, and \( MAX_I \) is the maximum pixel value, set to 255.

\section{Evaluation of Baseline Similarity Functions for Transfer Learning}\label{exist}
We conduct baseline evaluations to assess the performance of (i) Euclidean Distance, (ii) Pearson Correlation, (iii) Maximum Mean Discrepancy (MMD), (iv) Central Moment Discrepancy (CMD), (v) Correlation Alignment (CORAL), (vi) Dynamic Time Warping (DTW), (vii) Time Warp Edit Distance 
(TWED) and (viii) Wasserstein Distance for transfer learning. We perform two sets of baseline analyses: single-source and multi-source transfer learning. In single-source transfer learning, the model is trained on a single source dataset before being fine-tuned on the target dataset \cite{he2023instance}. We compare the effectiveness of similarity functions when the source dataset is from the same industry (banking) versus different industries in predicting bank stock prices. In multi-source transfer learning, the model is trained on multiple source datasets before fine-tuning on the target dataset \cite{lai2022coverage}. Three scenarios are considered: sources are selected either solely from banking, from other industries, or a combination of both. In each evaluation, the similarity functions are used to select the most similar source datasets for transfer learning. The setting of baseline evaluation can be seen in Table \ref{setting}.

\subsection{Dataset Description}
We use Hong Kong stock price series data, comprising daily closing prices from the beginning of 2022 to the end of 2023, retrieved via the yfinance Python module. The study forecasts the next day's closing price based on the previous 10 days' prices. The dataset is split into training and test sets in a 7:3 ratio. Before training, the time series are scaled using MinMaxScaler, with the feature range set from -1 to 1. As shown in Table \ref{table:stocks}, the stock datasets include bank stocks and stocks from other industries. The bank stocks represent the top 11 market capitalizations on HKEX, while the non-bank stocks are the top 10 constituents of the Hang Seng Index by market capitalization.

\begin{table*}[h!]
\scriptsize
\caption{Stock Datasets}
\centering
\begin{tabular}{|l|c|c|c|c|c|c|c|c|c|c|c|}
\hline
Bank Stocks    & \begin{tabular}[c]{@{}l@{}}0939.\\ HK\end{tabular} & \begin{tabular}[c]{@{}l@{}}3988.\\ HK\end{tabular} & \begin{tabular}[c]{@{}l@{}}3328.\\ HK\end{tabular} & \begin{tabular}[c]{@{}l@{}}0011.\\ HK\end{tabular} & \begin{tabular}[c]{@{}l@{}}1288.\\ HK\end{tabular} & \begin{tabular}[c]{@{}l@{}}1398.\\ HK\end{tabular} & \begin{tabular}[c]{@{}l@{}}2388.\\ HK\end{tabular} & \begin{tabular}[c]{@{}l@{}}2888.\\ HK\end{tabular} & \begin{tabular}[c]{@{}l@{}}3968.\\ HK\end{tabular} & \begin{tabular}[c]{@{}l@{}}1658.\\ HK\end{tabular} & \begin{tabular}[c]{@{}l@{}}0005.\\ HK\end{tabular} \\ \hline
Other Industry & \begin{tabular}[c]{@{}l@{}}0700.\\ HK\end{tabular} & \begin{tabular}[c]{@{}l@{}}0941.\\ HK\end{tabular} & \begin{tabular}[c]{@{}l@{}}1299.\\ HK\end{tabular} & \begin{tabular}[c]{@{}l@{}}3690.\\ HK\end{tabular} & \begin{tabular}[c]{@{}l@{}}1810.\\ HK\end{tabular} & \begin{tabular}[c]{@{}l@{}}9988.\\ HK\end{tabular} & \begin{tabular}[c]{@{}l@{}}0883.\\ HK\end{tabular} & \begin{tabular}[c]{@{}l@{}}9999.\\ HK\end{tabular} & \begin{tabular}[c]{@{}l@{}}9618.\\ HK\end{tabular} & \begin{tabular}[c]{@{}l@{}}0388.\\ HK\end{tabular} &                                                    \\ \hline
\end{tabular}
\label{table:stocks}
\end{table*}

\subsection{Network Architecture and Evaluation Metrics}
This paper use the network in \cite{he2023instance}. We focus on forecasting financial time series using two network architectures: Deep Neural Network (DNN) and Long Short-Term Memory (LSTM). The DNN used in our study consists of four layers, including two hidden layers with 256 nodes each and an output layer with one node. LSTM, suitable for sequence prediction tasks, is composed of multiple LSTM cells arranged in layers. Our LSTM network comprises two layers with 256 cells each, followed by a single-node output layer. The hyper-parameters of both networks are shown Table \ref{hyperparameters}.

\begin{table}[h!]
\caption{Hyper-parameters of Neural Network}
\scriptsize
\centering
\begin{tabular}{|l|c|c|}
\hline
Hyper-parameters        & Pre-train   & Fine-tune     \\ \hline
Epochs        & 100   & 100     \\ \hline
Batch size    & 200   & 200     \\ \hline
Optimizer     & Adam  & Adam    \\ \hline
Learning rate & 0.001 & 0.00001 \\ \hline
First moment  & 0.9   & 0.9     \\ \hline
Second moment & 0.999 & 0.999   \\ \hline
Loss function & MSE   & MSE     \\ \hline
\end{tabular}
\label{hyperparameters}
\end{table}
  
Model performance is evaluated using Mean Absolute Percentage Error (MAPE\footnote{We use nnMAPE and lstmMAPE to distinguish between the MAPE values obtained from the deep neural network (nn) and the LSTM models, respectively; similar distinctions are applied for other metrics like RMSE and R-squared (nnRMSE, lstmRMSE, nn$R^2$, and lstm$R^2$)}), Root Mean Squared Error (RMSE), and R-squared (R2), where lower MAPE and RMSE and higher R2 indicate better performance.

MAPE measures the accuracy of the model by calculating the percentage difference between the predicted values and the actual values. It provides an intuitive understanding of the prediction error in percentage terms, making it easier to interpret the model's accuracy.

\begin{equation}
\text{MAPE} = \frac{100\%}{n} \sum_{i=1}^{n} \left| \frac{y_i - \hat{y}_i}{y_i} \right|
\end{equation}

RMSE is used to measure the average magnitude of the prediction errors, with errors being squared before averaging. This metric gives more weight to large errors, which is useful when large deviations are particularly undesirable.

\begin{equation}
\text{RMSE} = \sqrt{\frac{\sum (y_i - \hat{y}_i)^2}{n}}
\end{equation}

R-squared (R2) is a statistical measure that represents the proportion of the variance in the dependent variable that is predictable from the independent variables. It indicates how well the model’s predictions approximate the actual data points, with values closer to 1 suggesting a better fit.

\begin{equation}
R^2 = 1 - \frac{\sum (y_i - \hat{y}_i)^2}{\sum (y_i - \bar{y}_i)^2}
\end{equation}
where $y_i$ is the actual value and $\hat{y}_i$ is the predicted value. $n$ is the prediction period.

\vspace{-1.5em}
\begin{table}[h!]
\scriptsize
\caption{Single Source and Multi-source Transfer Learning Baseline Evaluations and Experiments}
\centering
\begin{tabular}{|l|c|c|c|}
\hline
Evaluation\&Experiment   & Source ($S_A$)    & Source ($S_B$)    & Target \\ \hline
Evaluation\&Experiment 1 & Bank              & ---             & Bank   \\ \hline
Evaluation\&Experiment 2 & Other Industries & ---              & Bank   \\ \hline
Evaluation\&Experiment 3 & Bank             & Bank             & Bank   \\ \hline
Evaluation\&Experiment 4 & Other Industries & Other Industries & Bank   \\ \hline
Evaluation\&Experiment 5 & Bank             & Other Industries & Bank   \\ \hline
\end{tabular}
\label{setting}
\end{table}

\begin{table*}[htb]
\caption{Result of Baseline Evaluation 1}
\label{eva1}
\centering
\scriptsize
\begin{threeparttable}
\begin{tabular}{|l|c|c|c|c|c|c|c|}
\hline
\begin{tabular}[c]{@{}l@{}}Similarity\\ Functions\end{tabular}                                     & D                         & nnMAPE                   & nnRMSE                   & nn$R^2$                  & lstmMAPE                 & lstmRMSE                 & lstm$R^2$                \\ \hline
\begin{tabular}[c]{@{}l@{}}Euclidean, Pearson,\\ MMD, Coral, DTW,\\ TWED, Wasserstein\end{tabular} & 2888.HK & \textit{\textbf{0.9090}} & \textit{\textbf{0.6784}} & \textit{\textbf{0.8862}} & \textit{\textbf{1.4436}} & \textit{\textbf{1.1158}} & \textit{\textbf{0.6923}} \\ \hline
CMD                                                                                                & 2388.HK                   & 0.9093                   & 0.6792                   & 0.8860                   & 2.0817                   & 1.6230                   & 0.3489                   \\ \hline
\end{tabular}
\begin{tablenotes}
    \item[(a)] $D$ stands for source dataset for single source transfer learning.
    \item[(b)] Euclidean, Pearson, MMD, Coral, DTW, TWED, Wasserstein being in the same cell means that the source dataset selected by these similarity functions are the same, i.e 2888.HK.
\end{tablenotes}
\end{threeparttable}
\end{table*}

\begin{table*}[htb]
\centering
\scriptsize
\caption{Result of Baseline Evaluation 2}
\begin{tabular}{|l|c|c|c|c|c|c|c|}
\hline
\begin{tabular}[c]{@{}l@{}}Similarity\\ Functions\end{tabular}                                     & D       & nnMAPE                   & nnRMSE                   & nn$R^2$                  & lstmMAPE                 & lstmRMSE                 & lstm$R^2$                \\ \hline
\begin{tabular}[c]{@{}l@{}}Euclidean, Pearson,\\ MMD, Coral, DTW,\\ TWED, Wasserstein\end{tabular} & 0941.HK & 0.9104                   & 0.6808                   & 0.8854                   & 2.2173                   & 1.5995                   & 0.3677                   \\ \hline
CMD                                                                                                & 1299.HK & \textit{\textbf{0.8985}} & \textit{\textbf{0.6679}} & \textit{\textbf{0.8898}} & \textit{\textbf{1.0438}} & \textit{\textbf{0.7731}} & \textit{\textbf{0.8523}} \\ \hline
\end{tabular}
\end{table*}

\begin{table*}[htb]
\centering
\scriptsize
\caption{Result of Baseline Evaluation 3}
\begin{threeparttable}
\begin{tabular}{|l|c|c|c|c|c|c|c|c|}
\hline
\begin{tabular}[c]{@{}l@{}}Similarity\\ Functions\end{tabular}                    & $D_1$   & $D_2$   & nnMAPE                   & nnRMSE                   & nn$R^2$                  & lstmMAPE                 & lstmRMSE                 & lstm$R^2$                \\ \hline
\begin{tabular}[c]{@{}l@{}}Euclidean, MMD, Coral,\\ DTW, Wasserstein\end{tabular} & 2888.HK & 3968.HK & \textit{\textbf{0.8910}} & 0.6667                   & 0.8901                   & 0.9477                   & 0.7025                   & 0.8780                   \\ \hline
Pearson                                                                           & 2888.HK & 1288.HK & 0.9122                   & 0.6781                   & 0.8863                   & 1.0180                   & 0.7493                   & 0.8612                   \\ \hline
CMD                                                                               & 2388.HK & 1658.HK & 0.8980                   & \textit{\textbf{0.6650}} & \textit{\textbf{0.8907}} & \textit{\textbf{0.9274}} & \textit{\textbf{0.6939}} & \textit{\textbf{0.8810}} \\ \hline
TWED                                                                              & 2888.HK & 3988.HK & 0.9240                   & 0.6869                   & 0.8834                   & 0.9624                   & 0.7196                   & 0.8720                   \\ \hline
\end{tabular}
\begin{tablenotes}
    \item[(a)] $D_1$ and $D_2$ stand for the two source datasets for multi-source transfer learning.
    \item[(b)] Euclidean, MMD, Coral, DTW, Wasserstein being in the same cell means that the source datasets selected by these similarity functions are the same, i.e 2888.HK and 3968.HK.
\end{tablenotes}
\end{threeparttable}
\end{table*}

\begin{table*}[htb]
\centering
\scriptsize
\caption{Result of Baseline Evaluation 4}
\begin{tabular}{|l|c|c|c|c|c|c|c|c|}
\hline
\begin{tabular}[c]{@{}l@{}}Similarity\\ Functions\end{tabular}        & $D_1$   & $D_2$   & nnMAPE                   & nnRMSE                   & nn$R^2$                  & lstmMAPE                 & lstmRMSE                 & lstm$R^2$                \\ \hline
\begin{tabular}[c]{@{}l@{}}Euclidean, DTW,\\ Wasserstein\end{tabular} & 0941.HK & 1299.HK & 0.9017                   & 0.6704                   & 0.8889                   & 1.1433                   & 0.8624                   & 0.8162                   \\ \hline
Pearson, TWED                                                         & 0941.HK & 0883.HK & 0.9157                   & 0.6807                   & 0.8855                   & 1.1643                   & 0.8597                   & 0.8173                   \\ \hline
MMD                                                                   & 0941.HK & 9618.HK & \textit{\textbf{0.8928}} & \textit{\textbf{0.6654}} & \textit{\textbf{0.8906}} & \textit{\textbf{0.9520}} & \textit{\textbf{0.7058}} & \textit{\textbf{0.8769}} \\ \hline
CMD                                                                   & 1229.HK & 0883.HK & 0.9101                   & 0.6769                   & 0.8867                   & 0.9840                   & 0.7237                   & 0.8706                   \\ \hline
Coral                                                                 & 0941.HK & 1810.HK & 0.9110                   & 0.6741                   & 0.8877                   & 1.0257                   & 0.7616                   & 0.8566                   \\ \hline
\end{tabular}
\end{table*}

\vspace{-1.5em}
\begin{table*}[htb!]
\centering
\scriptsize
\caption{Result of Baseline Evaluation 5}
\label{eva5}
\begin{tabular}{|l|c|c|c|c|c|c|c|c|}
\hline
\begin{tabular}[c]{@{}l@{}}Similarity\\ Functions\end{tabular}                                     & $D_1$   & $D_2$   & nnMAPE                   & nnRMSE                   & nn$R^2$                  & lstmMAPE                 & lstmRMSE                 & lstm$R^2$                \\ \hline
\begin{tabular}[c]{@{}l@{}}Euclidean, Pearson,\\ MMD, Coral, DTW,\\ TWED, Wasserstein\end{tabular} & 2888.HK & 0941.HK & 0.9017                   & 0.6704                   & 0.8889                   & 1.1433                   & \textit{\textbf{0.8624}} & \textit{\textbf{0.8162}} \\ \hline
CMD                                                                                                & 2388.HK & 1299.HK & \textit{\textbf{0.9157}} & \textit{\textbf{0.6807}} & \textit{\textbf{0.8855}} & \textit{\textbf{1.1643}} & 0.8597                   & 0.8173                   \\ \hline
\end{tabular}
\end{table*}

\subsection{Summary Analysis for Baseline Similarity Functions}
Table \ref{eva1} to \ref{eva5} conclude the effectiveness of different similarity functions in transfer learning across various scenarios. In single-source transfer learning, CMD stands out as the best similarity function when the source is from other industries, while multiple functions, including Euclidean and Pearson, perform well for banking datasets. In multi-source transfer learning, CMD consistently excels, especially when sources are from the banking sector or a mix of industries. MMD is most effective for multi-source scenarios involving non-banking industries. This analysis underscores CMD's robustness and adaptability across different transfer learning contexts.

The similarity functions correspond to the lowest errors are considered the baseline best similarity functions. Table \ref{single_source} presents the baseline best similarity functions for single source transfer learning and Table \ref{multi_source} presents the baseline best similarity functions for multi-source transfer learning, providing a standard against which the GAF-based similarity functions will be compared in subsequent sections. 0005.HK is used as the target dataset, i.e $D_t=0005.HK$, and also for the experiments in Section \ref{experiment}.

\begin{table}[h!]
\caption{Baseline Best Similarity Functions for Single Source Transfer Learning}
\scriptsize
\centering
\begin{tabular}{|c|c|c|}
\hline
                                  & DNN                                                                                                & LSTM                                                                                               \\ \hline
($S_A$: Bank, T: Bank)             & \begin{tabular}[c]{@{}l@{}}Euclidean, Pearson,\\ MMD, Coral, DTW,\\ TWED, Wasserstein\end{tabular} & \begin{tabular}[c]{@{}l@{}}Euclidean, Pearson,\\ MMD, Coral, DTW,\\ TWED, Wasserstein\end{tabular} \\ \hline
($S_A$: Other Industries, T: Bank) & CMD                                                                                                & CMD                                                                                                \\ \hline
\end{tabular}
\label{single_source}
\end{table}

\begin{table}[h!]
\caption{Baseline Best Similarity Functions for Multisource Transfer Learning}
\scriptsize
\centering
\begin{tabular}{|p{4.5cm}|l|l|}
\hline
                                                          & DNN & LSTM                                                                                               \\ \hline
($S_A$: Bank, $S_B$: Bank, T: Bank)                         & CMD & CMD                                                                                                \\ \hline
($S_A$: Other Industries, $S_B$: Other Industries, T: Bank) & MMD & MMD                                                                                                \\ \hline
($S_A$: Bank, $S_B$: Other Industries, T: Bank)             & CMD & \begin{tabular}[c]{@{}l@{}}Euclidean, Pearson,\\ MMD, Coral, DTW,\\ TWED, Wasserstein\end{tabular} \\ \hline
\end{tabular}
\label{multi_source}
\end{table}

\section{Experiment}\label{experiment}
To evaluate the GAF-based similarity functions, we repeat the baseline evaluations from Section \ref{exist} using these GAF similarity functions. The experiments are denoted as experiments 1-5 as shown in Table \ref{setting}, corresponding to the previous baseline evaluations.

\subsection{Experiment Results}
The results of these experiments are shown in Table \ref{result1} to Table \ref{result5}. The errors of the GAF-based similarity functions are compared with the baseline best similarity functions' results in Tables \ref{single_source} and \ref{multi_source}. If the error obtained by GAF-based similarity functions is lower than that of similarity functions without GAF scaling, then GAF enhances the model's performance. The similarity functions corresponding to the lowest errors are considered the best similarity functions.


\begin{table*}[h!]
\caption{Result of Experiment 1}
\label{result1}
\centering
\scriptsize
\begin{tabular}{|lccccccc|}
\hline
\multicolumn{1}{|l|}{\begin{tabular}[c]{@{}l@{}}Similarity Functions\\ with GAF\end{tabular}} & \multicolumn{1}{c|}{D} & \multicolumn{1}{c|}{nnMAPE}                   & \multicolumn{1}{c|}{nnRMSE}                   & \multicolumn{1}{c|}{nn$R^2$}                  & \multicolumn{1}{c|}{lstmMAPE}                 & \multicolumn{1}{c|}{lstmRMSE}                 & lstm$R^2$                \\ \hline
\multicolumn{1}{|l|}{\begin{tabular}[c]{@{}l@{}}Euclidean, ARE,\\ PSNR, SSIM\end{tabular}}    & \multicolumn{1}{c|}{2888.HK}        & \multicolumn{1}{c|}{0.9090}                   & \multicolumn{1}{c|}{0.6784}                   & \multicolumn{1}{c|}{0.8862}                   & \multicolumn{1}{c|}{1.4436}                   & \multicolumn{1}{c|}{1.1158}                   & 0.6923                   \\ \hline
\multicolumn{1}{|l|}{CMD}                                                                     & \multicolumn{1}{c|}{3988.HK}        & \multicolumn{1}{c|}{0.9166}                   & \multicolumn{1}{c|}{0.6824}                   & \multicolumn{1}{c|}{0.8849}                   & \multicolumn{1}{c|}{\textit{\textbf{1.1058}}} & \multicolumn{1}{c|}{\textit{\textbf{0.8169}}} & \textit{\textbf{0.8350}} \\ \hline
\multicolumn{1}{|l|}{Coral}                                                                   & \multicolumn{1}{c|}{0011.HK}        & \multicolumn{1}{c|}{\textit{\textbf{0.8984}}} & \multicolumn{1}{c|}{\textit{\textbf{0.6708}}} & \multicolumn{1}{c|}{\textit{\textbf{0.8888}}} & \multicolumn{1}{c|}{1.7527}                   & \multicolumn{1}{c|}{1.3123}                   & 0.5744                   \\ \hline
\multicolumn{1}{|l|}{Wasserstein}                                                             & \multicolumn{1}{c|}{1288.HK}        & \multicolumn{1}{c|}{0.9152}                   & \multicolumn{1}{c|}{0.6800}                   & \multicolumn{1}{c|}{0.8857}                   & \multicolumn{1}{c|}{1.1467}                   & \multicolumn{1}{c|}{0.8497}                   & 0.8215                   \\ \hline
\multicolumn{8}{|l|}{}                                                                                                                                                                                                                                                                                                                                                                                         \\ \hline
\multicolumn{1}{|l|}{Best result without GAF}                                                 & \multicolumn{1}{c|}{2888.HK}        & \multicolumn{1}{c|}{0.9090}                   & \multicolumn{1}{c|}{0.6784}                   & \multicolumn{1}{c|}{0.8862}                   & \multicolumn{1}{c|}{1.4436}                   & \multicolumn{1}{c|}{1.1158}                   & 0.6923                   \\ \hline
\end{tabular}
\label{result1}
\end{table*}


\begin{table*}[h!]
\caption{Result of Experiment 2}
\centering
\scriptsize
\begin{tabular}{|lccccccc|}
\hline
\multicolumn{1}{|l|}{\begin{tabular}[c]{@{}l@{}}Similarity Functions\\ with GAF\end{tabular}} & \multicolumn{1}{c|}{D} & \multicolumn{1}{c|}{nnMAPE}                    & \multicolumn{1}{c|}{nnRMSE}                   & \multicolumn{1}{c|}{nn$R^2$}                  & \multicolumn{1}{c|}{lstmMAPE}                 & \multicolumn{1}{c|}{lstmRMSE}                 & lstm$R^2$                \\ \hline
\multicolumn{1}{|l|}{\begin{tabular}[c]{@{}l@{}}Euclidean, CMD,\\ Wasserstein\end{tabular}}   & \multicolumn{1}{c|}{9999.HK}        & \multicolumn{1}{c|}{0.9234}                    & \multicolumn{1}{c|}{0.6815}                   & \multicolumn{1}{c|}{0.8852}                   & \multicolumn{1}{c|}{1.1029}                   & \multicolumn{1}{c|}{0.8153}                   & 0.8357                   \\ \hline
\multicolumn{1}{|l|}{Coral}                                                                   & \multicolumn{1}{c|}{1299.HK}        & \multicolumn{1}{c|}{\textit{\textbf{0.8985}}} & \multicolumn{1}{c|}{\textit{\textbf{0.6679}}} & \multicolumn{1}{c|}{\textit{\textbf{0.8898}}} & \multicolumn{1}{c|}{\textit{\textbf{1.0438}}} & \multicolumn{1}{c|}{\textit{\textbf{0.7731}}} & \textit{\textbf{0.8523}} \\ \hline
\multicolumn{1}{|l|}{\begin{tabular}[c]{@{}l@{}}ARE, PSNR,\\ SSIM\end{tabular}}               & \multicolumn{1}{c|}{0883.HK}        & \multicolumn{1}{c|}{0.9128}                    & \multicolumn{1}{c|}{0.6787}                   & \multicolumn{1}{c|}{0.8862}                   & \multicolumn{1}{c|}{1.2799}                   & \multicolumn{1}{c|}{0.9424}                   & 0.7805                   \\ \hline
\multicolumn{8}{|l|}{}                                                                                                                                                                                                                                                                                                                                                                                          \\ \hline
\multicolumn{1}{|l|}{Best result without GAF}                                                 & \multicolumn{1}{c|}{1299.HK}        & \multicolumn{1}{c|}{\textbf{\textit{0.8985}}}                    & \multicolumn{1}{c|}{\textbf{\textit{0.6679}}}                   & \multicolumn{1}{c|}{\textbf{\textit{0.8898}}}                   & \multicolumn{1}{c|}{\textbf{\textit{1.0438}}}                   & \multicolumn{1}{c|}{\textbf{\textit{0.7731}}}                   & \textbf{\textit{0.8523}}                  \\ \hline
\end{tabular}
\label{result2}
\end{table*}

\begin{table*}[h!]
\caption{Result of Experiment 3}
\centering
\scriptsize
\begin{tabular}{|lllllllll|}
\hline
\multicolumn{1}{|l|}{\begin{tabular}[c]{@{}l@{}}Similarity Functions\\ with GAF\end{tabular}}                     & \multicolumn{1}{c|}{$D_1$}      & \multicolumn{1}{c|}{$D_2$}      & \multicolumn{1}{c|}{nnMAPE}                   & \multicolumn{1}{c|}{nnRMSE}                   & \multicolumn{1}{c|}{nn$R^2$}                    & \multicolumn{1}{c|}{lstmMAPE}                 & \multicolumn{1}{c|}{lstmRMSE}                 & lstm$R^2$                   \\ \hline
\multicolumn{1}{|l|}{\begin{tabular}[c]{@{}l@{}}Euclidean, \\ Wasserstein, \\ ARE, PSNR, \\ SSIM\end{tabular}} & \multicolumn{1}{l|}{2888.HK} & \multicolumn{1}{l|}{1288.HK} & \multicolumn{1}{c|}{0.9122}                   & \multicolumn{1}{c|}{0.6781}                   & \multicolumn{1}{c|}{0.8863}                   & \multicolumn{1}{c|}{1.0180}                   & \multicolumn{1}{c|}{0.7493}                   & \multicolumn{1}{c|}{0.8612}                   \\ \hline
\multicolumn{1}{|l|}{CMD}                                                                                      & \multicolumn{1}{l|}{3988.HK} & \multicolumn{1}{l|}{3328.HK} & \multicolumn{1}{c|}{0.9096}                   & \multicolumn{1}{c|}{0.6752}                   & \multicolumn{1}{c|}{0.8873}                   & \multicolumn{1}{c|}{1.0194}                   & \multicolumn{1}{c|}{0.7554}                   & \multicolumn{1}{c|}{0.8590}                   \\ \hline
\multicolumn{1}{|l|}{Coral}                                                                                    & \multicolumn{1}{l|}{0011.HK} & \multicolumn{1}{l|}{3988.HK} & \multicolumn{1}{c|}{0.9096}                   & \multicolumn{1}{c|}{0.6758}                   & \multicolumn{1}{c|}{0.8871}                   & \multicolumn{1}{c|}{0.9972}                   & \multicolumn{1}{c|}{0.7388}                   & \multicolumn{1}{c|}{0.8651}                   \\ \hline
\multicolumn{1}{|l|}{Wasserstein}                                                                              & \multicolumn{1}{l|}{1288.HK} & \multicolumn{1}{l|}{2888.HK} & \multicolumn{1}{c|}{0.9146}                   & \multicolumn{1}{c|}{0.6825}                   & \multicolumn{1}{c|}{0.8849}                   & \multicolumn{1}{c|}{1.3271}                   & \multicolumn{1}{c|}{1.0161}                   & \multicolumn{1}{c|}{0.7448}                   \\ \hline
\multicolumn{9}{|l|}{}                                                                                                                                                                                                                                                                                                                                                                                                                                  \\ \hline
\multicolumn{1}{|l|}{\multirow{2}{*}{\begin{tabular}[c]{@{}l@{}}Best Result\\ without GAF\end{tabular}}}       & \multicolumn{1}{l|}{2888.HK} & \multicolumn{1}{l|}{3968.HK} & \multicolumn{1}{c|}{\textit{\textbf{0.8910}}} & \multicolumn{1}{c|}{-} & \multicolumn{1}{c|}{-}                        & \multicolumn{1}{c|}{-}                        & \multicolumn{1}{c|}{-}                        & \multicolumn{1}{c|}{-}                       \\ \cline{2-9} 
\multicolumn{1}{|l|}{}                                                                                         & \multicolumn{1}{l|}{2388.HK} & \multicolumn{1}{l|}{1658.HK} & \multicolumn{1}{c|}{-}                        & \multicolumn{1}{c|}{\textit{\textbf{0.6650}}}                        & \multicolumn{1}{c|}{\textit{\textbf{0.8907}}} & \multicolumn{1}{c|}{\textit{\textbf{0.9274}}} & \multicolumn{1}{c|}{\textit{\textbf{0.6939}}} & \multicolumn{1}{c|}{\textit{\textbf{0.8810}}} \\ \hline
\end{tabular}
\label{result3}
\end{table*}

\begin{table*}[h!]
\caption{Result of Experiment 4}
\centering
\scriptsize
\begin{tabular}{|lcccccccc|}
\hline
\multicolumn{1}{|l|}{\begin{tabular}[c]{@{}l@{}}Similarity Functions\\ with GAF\end{tabular}}  & \multicolumn{1}{c|}{$D_1$}      & \multicolumn{1}{c|}{$D_2$}      & \multicolumn{1}{c|}{nnMAPE}                   & \multicolumn{1}{c|}{nnRMSE}                   & \multicolumn{1}{c|}{nn$R^2$}                  & \multicolumn{1}{c|}{lstmMAPE}                 & \multicolumn{1}{c|}{lstmRMSE}                 & lstm$R^2$                \\ \hline
\multicolumn{1}{|l|}{\begin{tabular}[c]{@{}l@{}}Euclidean,\\ Wasserstein,\\ PSNR\end{tabular}} & \multicolumn{1}{c|}{9999.HK} & \multicolumn{1}{c|}{0883.HK} & \multicolumn{1}{c|}{0.9201}                   & \multicolumn{1}{c|}{0.6838}                   & \multicolumn{1}{c|}{0.8844}                   & \multicolumn{1}{c|}{0.9760}                   & \multicolumn{1}{c|}{0.7193}                   & 0.8721                   \\ \hline
\multicolumn{1}{|l|}{CMD}                                                                      & \multicolumn{1}{c|}{9999.HK} & \multicolumn{1}{c|}{1299.HK} & \multicolumn{1}{c|}{0.9071}                   & \multicolumn{1}{c|}{0.6741}                   & \multicolumn{1}{c|}{0.8877}                   & \multicolumn{1}{c|}{1.0747}                   & \multicolumn{1}{c|}{0.8001}                   & 0.8418                   \\ \hline
\multicolumn{1}{|l|}{Coral}                                                                    & \multicolumn{1}{c|}{1299.HK} & \multicolumn{1}{c|}{9999.HK} & \multicolumn{1}{c|}{0.9203}                   & \multicolumn{1}{c|}{0.6792}                   & \multicolumn{1}{c|}{0.8860}                   & \multicolumn{1}{c|}{1.1314}                   & \multicolumn{1}{c|}{0.8416}                   & 0.8249                   \\ \hline
\multicolumn{1}{|l|}{ARE}                                                                      & \multicolumn{1}{c|}{0883.HK} & \multicolumn{1}{c|}{3690.HK} & \multicolumn{1}{c|}{0.8947}                   & \multicolumn{1}{c|}{0.6655}                   & \multicolumn{1}{c|}{0.8905}                   & \multicolumn{1}{c|}{\textit{\textbf{0.9184}}} & \multicolumn{1}{c|}{\textit{\textbf{0.6841}}} & \textit{\textbf{0.8843}} \\ \hline
\multicolumn{1}{|l|}{SSIM}                                                                     & \multicolumn{1}{c|}{0883.HK} & \multicolumn{1}{c|}{9999.HK} & \multicolumn{1}{c|}{0.9216}                   & \multicolumn{1}{c|}{0.6803}                   & \multicolumn{1}{c|}{0.8856}                   & \multicolumn{1}{c|}{1.0749}                   & \multicolumn{1}{c|}{0.7952}                   & 0.8437                   \\ \hline
\multicolumn{9}{|l|}{}                                                                                                                                                                                                                                                                                                                                                                                                                  \\ \hline
\multicolumn{1}{|l|}{\begin{tabular}[c]{@{}l@{}}Best Result\\ without GAF\end{tabular}}        & \multicolumn{1}{c|}{0941.HK} & \multicolumn{1}{c|}{0700.HK} & \multicolumn{1}{c|}{\textit{\textbf{0.8928}}} & \multicolumn{1}{c|}{\textit{\textbf{0.6654}}} & \multicolumn{1}{c|}{\textit{\textbf{0.8906}}} & \multicolumn{1}{c|}{0.9520}                   & \multicolumn{1}{c|}{0.7058}                   & 0.8769                   \\ \hline
\end{tabular}
\label{result4}
\end{table*}

\begin{table*}[h!]
\caption{Result of Experiment 5}
\centering
\scriptsize
\begin{tabular}{|lcccccccc|}
\hline
\multicolumn{1}{|l|}{\begin{tabular}[c]{@{}l@{}}Similarity Functions\\ with GAF\end{tabular}}         & \multicolumn{1}{c|}{$D_1$}      & \multicolumn{1}{c|}{$D_2$}      & \multicolumn{1}{c|}{nnMAPE}                   & \multicolumn{1}{c|}{nnRMSE}                   & \multicolumn{1}{c|}{nn$R^2$}                  & \multicolumn{1}{c|}{lstmMAPE}                 & \multicolumn{1}{c|}{lstmRMSE}                 & lstm$R^2$                \\ \hline
\multicolumn{1}{|l|}{\begin{tabular}[c]{@{}l@{}}Euclidean,  \\ PSNR\end{tabular}}                        & \multicolumn{1}{c|}{2888.HK} & \multicolumn{1}{c|}{9999.HK} & \multicolumn{1}{c|}{0.9289}                   & \multicolumn{1}{c|}{0.6848}                   & \multicolumn{1}{c|}{0.8841}                   & \multicolumn{1}{c|}{1.2318}                   & \multicolumn{1}{c|}{0.9340}                   & 0.7844                   \\ \hline
\multicolumn{1}{|l|}{CMD}                                                                                & \multicolumn{1}{c|}{3988.HK} & \multicolumn{1}{c|}{9999.HK} & \multicolumn{1}{c|}{0.9294}                   & \multicolumn{1}{c|}{0.6859}                   & \multicolumn{1}{c|}{0.8837}                   & \multicolumn{1}{c|}{1.0670}                   & \multicolumn{1}{c|}{\textit{\textbf{0.7849}}} & \textit{\textbf{0.8477}} \\ \hline
\multicolumn{1}{|l|}{Coral}                                                                              & \multicolumn{1}{c|}{0011.HK} & \multicolumn{1}{c|}{1299.HK} & \multicolumn{1}{c|}{\textit{\textbf{0.8930}}} & \multicolumn{1}{c|}{\textit{\textbf{0.6644}}} & \multicolumn{1}{c|}{\textit{\textbf{0.8909}}} & \multicolumn{1}{c|}{1.2119}                   & \multicolumn{1}{c|}{0.9197}                   & 0.7910                   \\ \hline
\multicolumn{1}{|l|}{Wasserstein}                                                                        & \multicolumn{1}{c|}{1288.HK} & \multicolumn{1}{c|}{9999.HK} & \multicolumn{1}{c|}{0.9294}                   & \multicolumn{1}{c|}{0.6858}                   & \multicolumn{1}{c|}{0.8837}                   & \multicolumn{1}{c|}{1.1039}                   & \multicolumn{1}{c|}{0.8224}                   & 0.8329                   \\ \hline
\multicolumn{1}{|l|}{ARE, SSIM}                                                                          & \multicolumn{1}{c|}{2888.HK} & \multicolumn{1}{c|}{0883.HK} & \multicolumn{1}{c|}{0.9164}                   & \multicolumn{1}{c|}{0.6807}                   & \multicolumn{1}{c|}{0.8855}                   & \multicolumn{1}{c|}{1.0774}                   & \multicolumn{1}{c|}{0.7950}                   & 0.8438                   \\ \hline
\multicolumn{9}{|l|}{}                                                                                                                                                                                                                                                                                                                                                                                                                            \\ \hline
\multicolumn{1}{|l|}{\multirow{2}{*}{\begin{tabular}[c]{@{}l@{}}Best Result\\ without GAF\end{tabular}}} & \multicolumn{1}{c|}{2388.HK} & \multicolumn{1}{c|}{1299.HK} & \multicolumn{1}{c|}{0.8942}                   & \multicolumn{1}{c|}{0.6648}                   & \multicolumn{1}{c|}{0.8908}                   & \multicolumn{1}{c|}{\textit{\textbf{1.0497}}} & \multicolumn{1}{c|}{-}                        & -                        \\ \cline{2-9} 
\multicolumn{1}{|l|}{}                                                                                   & \multicolumn{1}{c|}{2388.HK} & \multicolumn{1}{c|}{0941.HK} & \multicolumn{1}{c|}{-}                        & \multicolumn{1}{c|}{-}                        & \multicolumn{1}{c|}{-}                        & \multicolumn{1}{c|}{-}                        & \multicolumn{1}{c|}{0.7858}                   & 0.8474                   \\ \hline
\end{tabular}
\label{result5}
\end{table*}

\subsection{Summary Analysis for GAF-based Similarity Functions}
Table \ref{GAF_sum} combines the results of Section \ref{exist} and the experiment results, showcasing the best similarity functions under different conditions. The analysis highlights that scaling the time series by GAF and applying similarity functions can significantly improve transfer learning outcomes in several scenarios. Here, we discuss the findings in more detail:

\textbf{Single Source Transfer Learning (Banking):}
When both the source and target datasets consist of banking stocks, applying GAF-based similarity functions leads to notable improvements. Specifically, Coral (GAF) for DNN and CMD (GAF) for LSTM yielded the best performance. This indicates that the temporal dependencies and correlations captured by GAF transformations enhance the model's ability to generalize from the source to the target domain within the same industry. This is particularly beneficial in financial sectors where capturing intricate temporal patterns is crucial.

\textbf{Single Source Transfer Learning (Other Industries):} In single source transfer learning scenarios with source data from other industries, both GAF-based and traditional similarity functions showed advantages for DNN and LSTM models. Coral (GAF) and CMD were the best performers for both models, indicating that GAF transformations improved temporal dependency capture, but CMD without GAF also yielded strong results. This suggests that while GAF-based methods enhance model generalization, traditional CMD remains equally effective. This highlights the utility of GAF transformations in preserving temporal structures and validates the robustness of traditional CMD in such scenarios.

\textbf{Multi-source Transfer Learning (Banking):} In multi-source transfer learning scenarios where the source data is from banking industries, the results demonstrated that baseline similarity functions without GAF transformations provided the best performance for both DNN and LSTM models. This suggests that while GAF-based methods can enhance performance in other contexts, baseline similarity functions are more robust and effective in multi-source transfer learning when dealing with banking data.

\textbf{Multi-source Transfer Learning (Other Industries):}
For multi-source transfer learning using DNN and LSTM models, when the sources are from other industries, the GAF-based similarity functions demonstrated mixed results. Specifically, ARE (GAF) proved to be particularly effective for the LSTM model, showing significant improvement in capturing temporal structures and enhancing the model's performance. However, the DNN models did not benefit as much from the GAF transformations and did not exhibit a notable improvement compared to traditional methods. This suggests that while GAF transformations can effectively enhance LSTM models by preserving important temporal structures, their benefits may not extend to DNN models in multi-source transfer learning scenarios involving diverse industries.

\textbf{Multi-source Transfer Learning (Banking and Other Industries):}
In multi-source transfer learning scenarios where the first source is from banking industries and the second source is from other industries, both Coral (GAF) for DNN and CMD (GAF) for LSTM showed the best results. This finding is significant because it indicates that GAF transformations facilitate effective knowledge transfer even when the source domains are heterogeneous. The ability to integrate information from both similar and dissimilar sources can lead to more robust models, particularly in complex financial environments where different sectors may exhibit interconnected dynamics.

\vspace{-1em}

\begin{table}[htp!]
\scriptsize
\centering
\caption{Summary of Results}
\label{GAF_sum}
\begin{tabular}{|l|c|c|}
\hline
                                                                                                      & DNN                                                                                        & LSTM               \\ \hline
($S_A$: Bank, T: Bank)                                                                                & \textbf{Coral (GAF)}                                                                       & \textbf{CMD (GAF)} \\ \hline
\begin{tabular}[c]{@{}l@{}}($S_A$: Other Industries, \\ T: Bank)\end{tabular}                         & Coral (GAF), CMD                                                                           & Coral (GAF), CMD   \\ \hline
\begin{tabular}[c]{@{}l@{}}($S_A$: Bank, $S_B$: Bank,\\ T: Bank)\end{tabular}                         & \begin{tabular}[c]{@{}c@{}}CMD\end{tabular} & CMD                \\ \hline
\begin{tabular}[c]{@{}l@{}}($S_A$: Other Industries, $S_B$:\\ Other Industries, T: Bank)\end{tabular} & MMD                                                                                        & \textbf{ARE (GAF)} \\ \hline
\begin{tabular}[c]{@{}l@{}}($S_A$: Bank, $S_B$: \\ Other Industries, T: Bank)\end{tabular}            & \textbf{Coral (GAF)}                                                                       & \textbf{CMD (GAF)} \\ \hline
\end{tabular}
\end{table}

\subsection{Discussion}
In this subsection, we provide a deeper analysis of why GAF performs better. Discussing the underlying reasons for the performance improvements would provide more insights into the effectiveness of GAF in transfer learning, particularly in the context of financial time series data.

\subsubsection{Enhanced Temporal Feature Preservation} 
One of the key advantages of GAF-based similarity functions is their ability to capture and preserve temporal dependencies in time series data, which traditional methods like Euclidean distance or Pearson correlation often overlook. Traditional methods treat time points as isolated entities, missing the overall context and flow of the series. GAF transforms time series into 2D images, where each pixel encodes relationships between time points, maintaining the temporal structure and rhythm of the data. This richer representation is particularly valuable in financial markets, where the timing of price movements is critical to predicting future trends and patterns.

\subsubsection{Robustness to Noise and Nonlinearities} 
Financial time series data are typically noisy and nonlinear, and traditional similarity functions often struggle to handle these characteristics effectively. GAF-based methods, however, excel in these scenarios by smoothing out high-frequency noise and focusing on underlying trends. The angular transformation used in GAF highlights the direction and strength of changes, making it easier to detect and work with nonlinear patterns in the data. This leads to more robust models that can generalize across different market conditions, a crucial advantage for financial forecasting, where volatility and unexpected shifts are common.

\subsubsection{Improved Interpretability and Visual Analysis} 
Another advantage of GAF-based similarity functions is the improved interpretability they offer by transforming time series into images. This visual representation allows for a more intuitive understanding and comparison of different time series, making it easier to identify patterns, anomalies, or correlations between datasets. This interpretability is particularly useful in multi-source transfer learning, where understanding relationships between domains is crucial. By visually inspecting GAF images, researchers can quickly assess the similarity of different time series, aiding in the selection of the most appropriate source domains for transfer learning.

\subsubsection{Synergistic Integration with Deep Learning Models} 
GAF-based methods integrate seamlessly with deep learning models, especially those designed for image analysis like convolutional neural networks (CNNs). When time series are transformed into GAF images, they can be processed similarly to other image data, allowing the deep learning model to leverage spatial patterns and dependencies that traditional similarity functions might miss. This synergy between GAF and deep learning architectures enables the extraction of both temporal and spatial features, leading to more accurate predictions and better overall performance. This is particularly beneficial in complex financial environments where subtle patterns are crucial for accurate forecasting.

\subsubsection{Flexibility in Handling Heterogeneous Data Sources} 
GAF-based methods also provide significant flexibility in handling data from diverse sources, an area where traditional similarity functions can struggle. In multi-source transfer learning, datasets often come from different industries or exhibit varying behaviors. GAF standardizes these datasets by transforming them into a common 2D representation, allowing for a more consistent comparison across domains. This flexibility is invaluable in financial applications, where datasets can vary widely in structure and behavior. By enabling consistent comparisons through GAF, transfer learning models can more effectively integrate heterogeneous data, leading to improved generalization and performance across different domains.

\section{Conclusion}
In this study, we evaluated and compared the performance of various baseline similarity functions and Gramian Angular Field (GAF)-based similarity functions in the context of transfer learning for financial time series. The results from the baseline evaluations indicated that the Central Moment Discrepancy (CMD) function consistently performed well across different scenarios, especially in minimizing prediction errors and enhancing the transfer learning process. This robustness was particularly evident in multi-source transfer learning scenarios, where CMD effectively integrated knowledge from multiple sources.

When introducing GAF-based similarity functions, the findings revealed that these functions significantly improved the performance of transfer learning in certain contexts, particularly for single-source transfer learning within the banking industry. Specifically, Coral (GAF) for Deep Neural Networks (DNN) and CMD (GAF) for Long Short-Term Memory (LSTM) networks consistently delivered superior results, highlighting the effectiveness of GAF transformations in capturing temporal dependencies and correlations. However, the effectiveness of GAF-based functions was less consistent in multi-source transfer learning scenarios, particularly when the source domains involved a mix of banking and other industries. While some GAF-based functions like Adapted Rand Error (ARE) showed significant improvements for LSTM models, the benefits were not as pronounced for DNN models. This suggests that while GAF transformations can enhance the transfer learning process, particularly in capturing complex temporal structures, their utility may vary depending on the specific model and transfer learning scenario.

Overall, the study underscores the importance of selecting appropriate similarity functions for transfer learning, particularly when dealing with complex financial time series data. The findings suggest that while traditional similarity functions like CMD remain robust across various scenarios, GAF-based functions offer significant advantages in certain contexts, particularly in enhancing model performance in single-source transfer learning within the same industry. These insights provide a foundation for further research into optimizing transfer learning methodologies in financial time series analysis.


\end{document}